\documentclass[conference]{IEEEtran}

\usepackage[utf8]{inputenc}

\usepackage{amsmath,amssymb,epsfig,psfrag,cite,subfigure}
\include{macros}
\usepackage{graphicx}
\usepackage{epstopdf}

\usepackage{acronym}

\usepackage{bm}

\usepackage{geometry}
\usepackage{tcolorbox}

\usepackage{accents}
\newcommand*{\dt}[1]{%
	\accentset{\mbox{\large .}}{#1}}

\allowdisplaybreaks

\usepackage{soul}

\usepackage{amsthm}

\usepackage{nicefrac}

\usepackage{lipsum,multicol}

\usepackage{algorithm}
\usepackage{algpseudocode}

\algdef{SE}[SUBALG]{Indent}{EndIndent}{}{\algorithmicend\ }%
\algtext*{Indent}
\algtext*{EndIndent}

\usepackage{enumitem}
\usepackage{mwe}    
\usepackage{epsfig,psfrag}
\usepackage{subfigure}
\usepackage{color}
\usepackage{url}
\usepackage{mathtools,xparse}

\usepackage{gensymb}

\usepackage[multiple]{footmisc}

\usepackage{xr}

\makeatletter
\newcommand*{\addFileDependency}[1]{
  \typeout{(#1)}
  \@addtofilelist{#1}
  \IfFileExists{#1}{}{\typeout{No file #1.}}
}
\makeatother

\newcommand*{\myexternaldocument}[1]{%
    \externaldocument{#1}%
    \addFileDependency{#1.tex}%
    \addFileDependency{#1.aux}%
}

\myexternaldocument{supp}

\usepackage{xcolor}

\theoremstyle{remark}

\newtheoremstyle{mytheoremstyle} 
    {\topsep}                    
    {\topsep}                    
    {\upshape}                   
    {.5em}                           
    {\itshape}                   
    {.}                          
    {.5em}                       
    {}  
    
\theoremstyle{mytheoremstyle}

\newtheoremstyle{iremark}
  {\topsep}   
  {\topsep}   
  {\upshape}  
  {0.2in}       
  {\itshape}  
  {.}         
  {5pt plus 1pt minus 1pt} 
  {\thmname{#1}\thmnumber{ \itshape#2}\thmnote{ (#3)}} 

\theoremstyle{iremark}

\DeclareFontFamily{U}{mathx}{\hyphenchar\font45}
\DeclareFontShape{U}{mathx}{m}{n}{
	<5> <6> <7> <8> <9> <10>
	<10.95> <12> <14.4> <17.28> <20.74> <24.88>
	mathx10
}{}
\DeclareSymbolFont{mathx}{U}{mathx}{m}{n}
\DeclareFontSubstitution{U}{mathx}{m}{n}

\DeclareMathOperator*{\E}{\mathbb{E}}

\DeclarePairedDelimiter\abs{\lvert}{\rvert}%

\makeatletter \renewcommand\d[1]{\ensuremath{%
		\;\mathrm{d}#1\@ifnextchar\d{\!}{}}}
\makeatother

\makeatletter
\newcommand*\rel@kern[1]{\kern#1\dimexpr\macc@kerna}
\newcommand*\widebar[1]{%
  \begingroup
  \def\mathaccent##1##2{%
    \rel@kern{0.8}%
    \overline{\rel@kern{-0.8}\macc@nucleus\rel@kern{0.2}}%
    \rel@kern{-0.2}%
  }%
  \macc@depth\@ne
  \let\math@bgroup\@empty \let\math@egroup\macc@set@skewchar
  \mathsurround\z@ \frozen@everymath{\mathgroup\macc@group\relax}%
  \macc@set@skewchar\relax
  \let\mathaccentV\macc@nested@a
  \macc@nested@a\relax111{#1}%
  \endgroup
}
\makeatother

\newcommand{\yy}{\mathbf{y}}

\newcommand{\ff}{\mathbf{f}}

\newcommand{\bb}{\mathbf{b}}
\newcommand{\zz}{\mathbf{z}}

\newcommand{\pp}{\mathbf{p}}
\newcommand{\qq}{\mathbf{q}}

\newcommand{\hermit}{H}
\newcommand{\trpose}{T}

\newcommand{\Imatrix}{{ \boldsymbol{\mathrm{I}} }}

\newcommand{\vecc}[1]{ {\rm{vec}}\left(#1\right)  }

\newcommand{\fb}{\mathbf{f}}

\newcommand{\alphare}[1]{ \alpha_{\rm{R}} }
\newcommand{\alphaim}[1]{ \alpha_{\rm{I}} }

\newcommand{\Psq}{ \sqrt{P} }

\newcommand{\thetaminl}{\theta_{{\rm{min}},\ell}}
\newcommand{\thetamaxl}{\theta_{{\rm{max}},\ell}}

\newcommand{\Ngrid}{N_{\rm{grid}}}

\newcommand{\BB}{\mathbf{B}}
\newcommand{\UU}{\mathbf{U}}
\newcommand{\QQ}{\mathbf{Q}}
\newcommand{\Lambdab}{\mathbf{\Lambda}}

\newcommand{\Lambdabhatls}{\widehat{\Lambdab}_{\rm{LS}}}

\newcommand{\FF}{\mathbf{F}}

\newcommand{\CC}{\mathbf{C}}

\newcommand{\snr}{{\rm{SNR}}}
\newcommand{\snrl}{\snr_{\ell}}

\newcommand{\nl}{ N_{\ell} }

\newcommand{\RMSE}{ {\rm{RMSE}} }

\usepackage{accents}

\newcommand{\FFsum}{\FF^{\rm{sum}}}
\newcommand{\FFdiff}{\FF^{\rm{diff}}}
\newcommand{\FFopt}{\FF^{\rm{base}}}

\newcommand{\JJ}{\mathbf{J}}
\newcommand{\TT}{\mathbf{T}}

\newcommand{\tracesmall}[1]{ {{{\rm{tr}}\left( #1 \right)}}  }

\newcommand{\AAb}{\mathbf{A}}
\newcommand{\RRb}{\mathbf{R}}
\newcommand{\VVb}{\mathbf{V}}

\newcommand{\atxdt}{\dt{\mathbf{a}} }
\newcommand{\eeb}{{\mathbf{e}} }

\newcommand{\norm}[1]{\left\lVert#1\right\rVert}
\newcommand{\norms}[1]{\big\lVert#1\big\rVert}

\newcommand{\rmtx}{{\rm{Tx}}}

\newcommand{\Ntx}{N_\rmtx}

\newcommand{\atx}{\mathbf{a}}

\newcommand{\alphal}{ \alpha_{\ell} }

\newcommand{\mtul}{ \mtu_{\ell} }

\newcommand{\realp}[1]{ \Re \left\{#1\right\}  }

\newcommand{\thetab}{ \bm{\theta} }

\newcommand{\thetabhat}{ \widehat{\thetab} }

\newcommand{\thetahat}{ \widehat{\theta} }

\newcommand{\alphab}{ \bm{\alpha} }

\newcommand{\fc}{ f_c }

\newcommand{\complexset}[2]{ \mathbb{C}^{#1 \times #2}  }

\newcommand{\realset}[2]{ \mathbb{R}^{#1 \times #2}  }

\newcommand{\thn}[1]{ {{#1^{\rm{th}}}} }

\newcommand{\mtCN}{{\mathcal{CN}}}
\newcommand{\mtu}{{\mathcal{U}}}
\newcommand{\boldzero}{{ {\boldsymbol{0}} }}

\newcommand{\conj}{{ * }}

\acrodef{RIS}{reconfigurable intelligent surface}
\acrodef{SNR}{signal-to-noise ratio}
\acrodef{ISAC}{integrated sensing and communication}
\acrodef{ISLAC}{integrated sensing, localization, and communication}
\acrodef{LoS}{line-of-sight}
\acrodef{NLoS}{non-line-of-sight}
\acrodef{AoA}{angle-of-arrival}
\acrodef{AoD}{angle-of-departure}
\acrodef{ToA}{time-of-arrival}
\acrodef{UE}{user equipment}
\acrodef{NF}{near-field}
\acrodef{BS}{base station}
\acrodef{MCRB}{misspecified Cram\'{e}r-Rao bound}
\acrodef{CRB}{Cram\'{e}r-Rao bound}
\acrodef{LB}{lower bound}
\acrodef{ML}{maximum-likelihood}
\acrodef{MML}{mismatched maximum-likelihood}
\acrodef{DL}{downlink}
\acrodef{UL}{uplink}
\acrodef{MIMO}{multiple-input multiple-output}
\acrodef{MISO}{multiple-input single-output}
\acrodef{SIP}{shift invariance property}
\acrodef{FIM}{Fisher information matrix}
\acrodef{RMSE}{root mean-squared error}
\acrodef{AWGN}{additive white Gaussian noise}
\acrodef{ADMM}{alternating direction method of multipliers}
\acrodef{LS}{least-squares}
\acrodef{SOC}{second-order cone}
\acrodef{SOCP}{second-order cone programming}
\acrodef{OFDM}{orthogonal frequency-division multiplexing}

\graphicspath{{./Figures/}}

\begin{document}
\bstctlcite{IEEEexample:BSTcontrol}

\title{ESPRIT-Oriented Precoder Design for mmWave Channel Estimation}

\author{Musa Furkan Keskin\IEEEauthorrefmark{1}, 
Alessio Fascista\IEEEauthorrefmark{2}, 
Fan Jiang\IEEEauthorrefmark{3}, 
Angelo Coluccia\IEEEauthorrefmark{2}, \\
Gonzalo Seco-Granados\IEEEauthorrefmark{4},
Henk Wymeersch\IEEEauthorrefmark{1}
\\
\IEEEauthorrefmark{1}Chalmers University of Technology, Sweden, 
\IEEEauthorrefmark{2}University of Salento, Italy, \\
\IEEEauthorrefmark{3}Halmstad University, Sweden,
\IEEEauthorrefmark{4}Autonomous University of Barcelona, Spain}

\maketitle

\begin{abstract}
We consider the problem of ESPRIT-oriented precoder design for beamspace angle-of-departure (AoD) estimation in downlink mmWave multiple-input single-output communications. Standard precoders (i.e., directional/sum beams) yield poor performance in AoD estimation, while Cram\'{e}r-Rao bound-optimized precoders undermine the so-called shift invariance property (SIP) of ESPRIT. To tackle this issue, the problem of designing ESPRIT-oriented precoders is formulated to jointly optimize over the precoding matrix and the SIP-restoring matrix of ESPRIT. We develop an alternating optimization approach that updates these two matrices under unit-modulus constraints for analog beamforming architectures. Simulation results demonstrate the validity of the proposed approach while providing valuable insights on the beampatterns of the ESPRIT-oriented precoders.
\end{abstract}

\begin{IEEEkeywords}
Precoder design, beamspace ESPRIT, channel estimation, mmWave communications.
\end{IEEEkeywords}

\section{Introduction}\label{sec_intro}

Positioning in 5G relies to a large extent on the use of mmWave frequencies, with their ample bandwidth and large antenna arrays \cite{b5g_commag_2021,dwivedi2021positioning,Fascista_ICASSP2020}. Large bandwidths offer high delay resolution but provide limited opportunities for optimization, as \acp{BS} must use non-overlapping subcarriers for multi-BS positioning solutions. Large antenna arrays yield high angle resolution, as well as the ability to shape signals in the spatial domain, e.g., for interference control, but also for optimizing positioning performance \cite{signalDesign_TVT_2022}. Harnessing the improved resolution and also exploiting optimized spatial designs enhance the performance of the channel estimation routine, which detects the number of paths, and for each path estimates the geometric parameters (i.e., \ac{ToA}, \ac{AoA}, \ac{AoD}) \cite{TR38.855}. As channel estimation is a joint function among communication, positioning, and sensing, it is important to develop methods that are both accurate and of moderate complexity \cite{Fascista_WCL}, especially for \ac{ISAC} systems towards 6G multi-functional wireless networks \cite{overviewISAC_2021}.

In a general pilot-based channel estimation setup, the optimal channel parameters are the maximum \emph{a posteriori} (MAP) estimates given the received signal sequence. However, optimization methods employed in MAP estimation can involve heavy computations. On the other hand, it is notable that mmWave channels are usually sparse, due to a limited number of multipath propagation arriving at the receiver with relatively strong path gains. As a result, sparsity-inspired low-complexity channel estimation methods are developed \cite{tsai2018millimeter, lee2016channel, alkhateeb2014channel, JiaGeZhuWym21, WenKulWitWym19, gridless_ESPRIT_JSTSP_2021}. Among them, the estimation of signal parameters via rotational invariance techniques (ESPRIT)-based channel estimation methods have been widely studied, due to their good trade-off between estimation performance and complexity \cite{JiaGeZhuWym21, WenKulWitWym19, gridless_ESPRIT_JSTSP_2021}. Recently, ESPRIT-based approaches have been applied to the beamspace, which is attractive since analogue and/or digital beamforming structures are employed in most massive MIMO mmWave systems \cite{gridless_ESPRIT_JSTSP_2021, Fan_ESPRIT_2021, WenGarKulWitWym18}. However, to apply beamspace ESPRIT methods, precoders are required to hold the \ac{SIP}.
Examples of such precoding matrices include the discrete Fourier transform (DFT) beams \cite{gridless_ESPRIT_JSTSP_2021, beamspace_ESPRIT} and the directional beams \cite{Fan_ESPRIT_2021}. When the \ac{SIP} does not hold for the precoding matrix, an approximation will be applied during the derivation of the beamspace ESPRIT methods, leading to performance degradations \cite{Fan_ESPRIT_2021}. In addition, research on \ac{CRB}-optimized precoder design suggests that the optimal precoding matrix usually does not hold the \ac{SIP} \cite{signalDesign_TVT_2022,precoderNil2018,Fascista_RIS}. In other words, there is an inevitable performance loss when low-complexity ESPRIT methods are employed with \ac{CRB}-optimized precoders.

\begin{figure}
	\centering
	\includegraphics[width=0.9\linewidth]{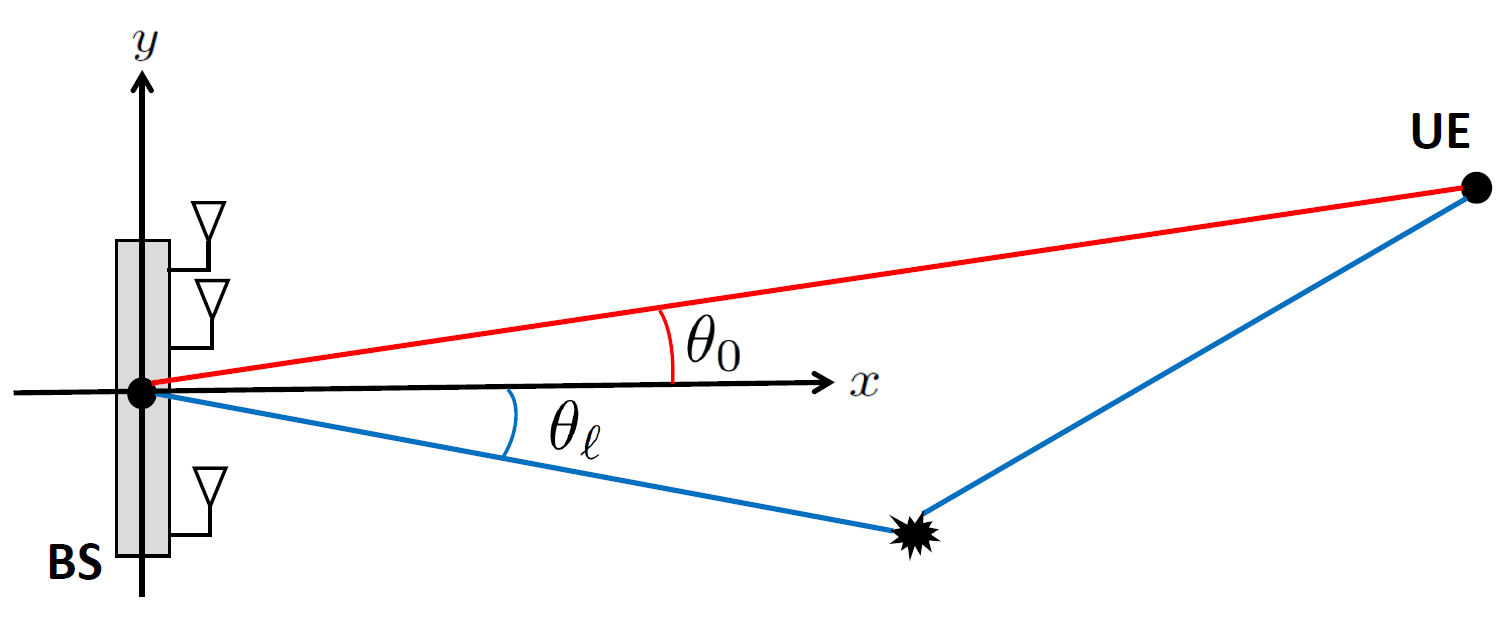}
	\caption{mmWave \ac{MISO} downlink scenario where the UE aims to estimate the \acp{AoD} of multiple paths using high-resolution beamspace ESPRIT methods.}
	\label{fig_scenario}
	\vspace{-0.2in}
\end{figure}

In this paper, we investigate the problem of ESPRIT-oriented precoder design for \ac{AoD} estimation in mmWave communications, targeting a near-optimal precoding scheme in terms of accuracy while enjoying the low-complexity and high-resolution ESPRIT methods for channel estimation. Our specific contributions are as follows:
\begin{itemize}
    \item We formulate the problem of ESPRIT-oriented precoder design as a beampattern synthesis problem that considers joint optimization of the precoding matrix and the \ac{SIP}-restoring matrix of ESPRIT.
    
    \item We propose an alternating optimization strategy that updates the precoder and the \ac{SIP}-restoring matrix sequentially under the unit-modulus constraint on individual precoder elements, suitable for phase-only beamforming architectures.
    
    \item Through simulation results, we provide important insights into the beampatterns of the resulting ESPRIT-oriented precoders and demonstrate the effectiveness of the proposed design approach in ESPRIT-based channel estimation.
\end{itemize}

\section{System Model and Problem Description} \label{sec_sysmod}

\subsection{System Model}
We consider a mmWave \ac{MISO} \ac{DL} flat-fading communications scenario with an $\Ntx$-antenna \ac{BS} and a single-antenna \ac{UE}, as shown in Fig.~\ref{fig_scenario}. Considering the presence of $L$ paths\footnote{We consider a mmWave tracking scenario \cite{precoderNil2018,mmWave_Tracking_2020_TVT,mmWave_Tracking_2020_TCOM} with known $L$.}, 
the received signal at the \ac{UE} at transmission instance $m$ and snapshot $n$ is given by
\begin{align}\label{eq_ym}
            y_{m,n} = \Psq \sum_{\ell=0}^{L-1} \alpha_{\ell,n} \, \atx^\mathsf{T}(\theta_{\ell}) \fb_m s_{m,n} + z_{m,n} 
\end{align}
for $m = 1, \ldots, M$ and $n = 1, \ldots, N$, where $M$ and $N$ denote, respectively, the number of transmissions and the number of snapshots\footnote{Here, snapshots may correspond to, for instance, different subcarriers of an \ac{OFDM} system. In this case, it is reasonable to assume that the channel gains $\alpha_{\ell,n}$ change across snapshots, but the \acp{AoD} $\theta_{\ell}$ remain constant.}. In \eqref{eq_ym}, $P$ denotes the transmit power, $[\atx(\theta)]_k = e^{j 2 \pi \frac{d}{\lambda} k \sin \theta}$, $k = 0 , \ldots , \Ntx-1$, is the steering vector for the \ac{BS} TX array, $\lambda = c/\fc$ is the wavelength with $c$ and $\fc$ denoting the speed of propagation and carrier frequency, respectively, $d$ is the array element spacing, $\fb_m \in \complexset{\Ntx}{1}$ denotes the \ac{BS} precoder at time $m$, $\alpha_{\ell,n}$ and $\theta_{\ell}$ are the complex channel gain and \ac{AoD} of the $\thn{\ell}$ path for the $\thn{n}$ snapshot, respectively, $s_{m,n}$ is the pilot symbol, and $z_{m,n} \sim \mtCN(0, \sigma^2)$ is \ac{AWGN} with power $\sigma^2$. For simplicity, we set $s_{m,n} = 1, \, \forall m, n$.

Aggregating the observations \eqref{eq_ym} over $M$ transmissions, we have the received signal at the $\thn{n}$ snapshot
\begin{align}\label{eq_yy}
    \yy_n = \Psq \FF^\mathsf{T} \VVb \alphab_n + \zz_n ~,
\end{align}
where $\yy_n \triangleq [y_{1,n} \,  \cdots \,  y_{M,n} ]^\mathsf{T} \in \complexset{M}{1}$, $\FF \triangleq \left[ \fb_1 \, \cdots \, \fb_M \right] \in \complexset{\Ntx}{M}$ is the precoding matrix satisfying $\tracesmall{\FF \FF^\mathsf{H}} = M$, $\VVb \triangleq \left[ \atx(\theta_0) \, \cdots \, \atx(\theta_{L-1}) \right] \in \complexset{\Ntx}{L}$, $\alphab_n \triangleq [\alpha_{0,n} \, \cdots \, \alpha_{L-1,n}]^\mathsf{T} \in \complexset{L}{1}$, and $\zz_n \sim \mtCN(\boldzero, \sigma^2 \Imatrix)$ represents the \ac{AWGN} component.

\subsection{Problem Description}
In the considered mmWave scenario, the \ac{UE} aims to estimate the \acp{AoD} $\thetab = [\theta_0 \, \cdots \, \theta_{L-1}]^\mathsf{T}$ using beamspace ESPRIT \cite{RoyKai89,beamspace_ESPRIT}
from the beamspace observations $\{\yy_n\}_{n=1}^{N}$ in \eqref{eq_yy}. The problem of interest is to design the \ac{BS} precoding matrix $\FF$ to maximize the accuracy of estimation of $\thetab$ at the UE while at the same time trying to preserve as much as possible the \ac{SIP} required by ESPRIT-based estimation \cite{beamspace_ESPRIT}.

\section{ESPRIT-Oriented Precoder Design} \label{sec_esprit}
In this section, we provide a review of beamspace ESPRIT and revisit the \ac{SIP}, which enforces a certain structure on the precoder. Based on this structure and using an ESPRIT-unaware baseline precoder $\FFopt$ (which will be introduced later in Sec.~\ref{sec_baseline}), we formulate a novel precoder design problem that jointly optimizes beampattern synthesis accuracy (with respect to $\FFopt$) and ESPRIT SIP error (i.e., the level of degradation of \ac{SIP}), leading to near-optimal performance for ESPRIT-based estimators.

\subsection{Review of Beamspace ESPRIT}\label{sec_beamspace_esprit}
From \eqref{eq_yy}, we compute the covariance matrix
\begin{align} \label{eq_rcov}
    \RRb = \frac{1}{N} \sum_{n=1}^N \yy_n \yy_n^\mathsf{H} = P \FF^\mathsf{T} \VVb \Big(\frac{1}{N} \sum_{n=1}^N \alphab_n \alphab_n^\mathsf{H} \Big) \VVb^\mathsf{H} \FF^* + \sigma^2 \Imatrix~,
\end{align}
where $\VVb$ is a Vandermonde matrix which holds the \ac{SIP}, satisfying $\JJ_1 \VVb = \JJ_2 \VVb \boldsymbol{\Phi}^\mathsf{H}$
where $\JJ_1 = \left[ \Imatrix_{\Ntx-1} ,~ \boldzero_{(\Ntx-1)\times 1} \right] \in \realset{(\Ntx-1)}{\Ntx}$ and $\JJ_2 = \left[ \boldzero_{(\Ntx-1)\times 1} , ~ \Imatrix_{\Ntx-1}  \right] \in \realset{(\Ntx-1)}{\Ntx}$ are selection matrices, and $\boldsymbol{\Phi} = \mathrm{Diag}([[\atx(\theta_0)]_1, [\atx(\theta_1)]_1, \cdots, [\atx(\theta_{L-1})]_1]^\mathsf{T})$. In \eqref{eq_rcov}, we assume that $\frac{1}{N} \sum_{n=1}^N \alphab_n \alphab_n^\mathsf{H}$ is a diagonal matrix (i.e., paths are decorrelated), meaning that the dimension of the signal subspace is $L$. In the precoded case, it has been shown in \cite{beamspace_ESPRIT, Fan_ESPRIT_2021} that if the matrix $\FF$ holds the \ac{SIP}, i.e.,
\begin{align}\label{eq_sip}
    \JJ_1 \FF = \JJ_2 \FF \Lambdab 
\end{align}
for some non-singular $\Lambdab \in \complexset{M}{M}$, we can restore the \ac{SIP} from $\CC = \FF^\mathsf{T} \VVb$, by finding a non-null matrix $\QQ$ such that
\begin{align}\label{eq_Bsip}
    \QQ \CC \boldsymbol{\Phi} = \QQ \Lambdab^\mathsf{T} \CC~,
\end{align}
where $\QQ \in \mathbb{C}^{M\times M}$ satisfies
\begin{align}\label{eq_Q}
    \QQ [\FF^{T}\eeb_M ~~ \Lambdab^\mathsf{T} \FF^\mathsf{T}\eeb_1 ]= \boldsymbol{0},
\end{align}
and $\eeb_m \in \realset{\Ntx}{1}$ is the $m$-th column of the identity matrix $\Imatrix_{\Ntx}$. From \eqref{eq_Q}, $\QQ$ can be obtained as $\QQ = \Imatrix_M - \sum_{i=0}^{1} \qq_i \qq_i^{\mathsf{H}} $, where $\qq_0, \qq_1 \in \complexset{M}{1}$ are orthonormal column vectors spanning the subspace corresponding to $[\FF^{T}\eeb_M ~~ \Lambdab^\mathsf{T} \FF^\mathsf{T}\eeb_1 ] \in \complexset{M}{2}$ \cite{Fan_ESPRIT_2021}. Since perfect SIP cannot always be guaranteed\footnote{Perfect SIP holds for DFT beams and directional/sum beams (i.e., steering vectors).} in \eqref{eq_sip}, one can resort to the least-squares (LS) solution to find an approximate $\Lambdab$ \cite{Fan_ESPRIT_2021}:
\begin{align}\label{eq_ls_sip_prob}
         \Lambdabhatls &= \mathrm{arg} \mathop{\mathrm{min}}\limits_{\Lambdab} ~ \norm{ \JJ_1 \FF - \JJ_2 \FF \Lambdab }_F^2  
         \\ \label{eq_ls_sip}
         &= \left( \FF^\hermit \JJ_2^\hermit \JJ_2 \FF  \right)^{-1} \FF^\hermit \JJ_2^\hermit \JJ_1 \FF ~,
\end{align}
where $\norm{\cdot}_F$ denotes the Frobenius norm. 

Given an estimate of the covariance matrix $\RRb$, the signal subspace matrix $\UU_\mathrm{s} \in \mathbb{C}^{M\times L}$ can be obtained through the SVD (or truncated SVD) operation. Since both $\CC$ and $\UU_\mathrm{s}$ span the same signal subspace, we have $\CC = \UU_\mathrm{s} \TT$,
where $\TT\in \mathbb{C}^{L\times L}$ is a non-singular matrix.
Using the \ac{SIP} of $\CC$ in \eqref{eq_Bsip}, we further obtain $\QQ \UU_{\mathrm{s}} \boldsymbol{\Pi} = \QQ\Lambdab^\mathsf{T}\UU_{\mathrm{s}}$ where $\boldsymbol{\Pi} = \TT \boldsymbol{\Phi} \TT^{-1}$. The diagonal elements in $\boldsymbol{\Phi}$ will be used to estimate the AoD of each path. The beamspace ESPRIT approach can be summarized as follows:
\begin{itemize}
    \item Find $\boldsymbol{\Lambda}$ and $\QQ$ for given $\FF$.
    \item Obtain an estimate of $\RRb$ as $\widetilde{\RRb}$ using multiple snapshots.
    \item Perform SVD (or truncated SVD) on $\widetilde{\RRb}$ to obtain $\UU_{\mathrm{s}}$.
    \item Obtain the least-square (LS) solution of $\boldsymbol{\Pi}$ as $\widetilde{\boldsymbol{\Pi}} = (\QQ\UU_{\mathrm{s}})^{\dagger} \QQ\Lambdab^\mathsf{T}\UU_{\mathrm{s}}$, where $(\cdot)^{\dagger}$ denotes Moore-Penrose pseudo-inverse.
    \item Perform eigenvalue decomposition on $\widetilde{\boldsymbol{\Pi}}$ to obtain an estimate of $\boldsymbol{\Phi}$, and retrieve the corresponding AoDs.
\end{itemize}

\subsection{ESPRIT-Oriented Precoder Design with SIP Considerations}

We formulate the problem of ESPRIT-oriented precoder design as a beampattern synthesis via joint optimization of $\FF$ and $\Lambdab$, starting from a desired beampattern created by an ESPRIT-unconstrained precoder $\FFopt$ as baseline. The goal is to minimize the weighted average of the beampattern synthesis error and the ESPRIT \ac{SIP} error, quantified by the error of the LS solution in \eqref{eq_ls_sip_prob}:
\begin{subequations} \label{eq_problem_sip_gen}
\begin{align} \label{eq_problem_sip}
	\mathop{\mathrm{min}}\limits_{\FF, \Lambdab} &~~ \underbrace{	\norm{ \BB - \AAb^\mathsf{T} \FF  }_F^2}_{\substack{{\rm{beampattern~synthesis}} \\ {\rm{accuracy}} }} + ~ \eta  \underbrace{ \norm{ \JJ_1 \FF - \JJ_2 \FF \Lambdab }_F^2 }_{\substack{{\rm{SIP\,approximation}} \\ {\rm{error}}  }}
	\\ \label{eq_problem_sip_cons_analog}
	~~ \mathrm{s.t.} &~~ \abs{ [\FF]_{n,m}} = 1 , \, \forall n,m ~,
\end{align}
\end{subequations}
where $\BB = \AAb^\mathsf{T} \FFopt  \in \complexset{\Ngrid}{M}$ represents the desired beampattern corresponding to $\FFopt$ at $\Ngrid$ angular grid points $\{ \theta_i \}_{i=1}^{\Ngrid}$, $\AAb = \left[ \atx(\theta_1) \, \cdots \, \atx(\theta_{\Ngrid}) \right] \in \complexset{\Ntx}{\Ngrid}$ is the transmit steering matrix evaluated at the specified grid locations, and $\eta$ is a predefined weight on the SIP error, chosen to provide a suitable trade-off between beampattern synthesis accuracy and SIP approximation error. In addition, the constraint \eqref{eq_problem_sip_cons_analog} is imposed to ensure compatibility with phase-only beamforming architectures \cite{analogBeamformerDesign_TSP_2017} (e.g., analog passive arrays \cite{phasedArray_2016}). In the case of phase-amplitude beamforming (e.g., via active phased arrays \cite{phasedArray_2016}), the problem becomes the special case of \eqref{eq_problem_sip_gen} without the constraint \eqref{eq_problem_sip_cons_analog}.

\subsection{Alternating Optimization to Solve \eqref{eq_problem_sip_gen}}
The problem \eqref{eq_problem_sip_gen} is non-convex due to \textit{(i)} the non-convexity of \eqref{eq_problem_sip} in the joint variable $\FF$ and $\Lambdab$, and \textit{(ii)} the unit-modulus constraint in \eqref{eq_problem_sip_cons_analog}. To tackle \eqref{eq_problem_sip_gen}, we resort to an alternating optimization method that updates $\FF$ and $\Lambdab$ in an iterative fashion. 


\subsubsection{Optimize $\FF$ for fixed $\Lambdab$}
Using the vectorization property of the Kronecker product, the objective function \eqref{eq_problem_sip} can be rewritten as
\begin{align}
    & g(\ff) \\ \nonumber &= \norm{ \bb - \left(\Imatrix_M \otimes \AAb^\trpose \right) \ff }_2^2 + \eta \norm{ \left(\Imatrix_M \otimes \JJ_1  -  \Lambdab^\trpose \otimes \JJ_2 \right) \ff }_2^2 ~,
\end{align}
where $\ff \triangleq \vecc{\FF}$ and $\bb \triangleq \vecc{\BB}$. Defining
\begin{align}
    \QQ &\triangleq \left(\Imatrix_M \otimes \AAb^\trpose \right)^\hermit \left(\Imatrix_M \otimes \AAb^\trpose \right) \\ \nonumber
    &~~+ \eta \left(\Imatrix_M \otimes \JJ_1  -  \Lambdab^\trpose \otimes \JJ_2 \right)^\hermit \left(\Imatrix_M \otimes \JJ_1  -  \Lambdab^\trpose \otimes \JJ_2 \right) ~,  \\ \nonumber
    \pp &\triangleq \left(\Imatrix_M \otimes \AAb^\trpose \right)^\hermit \bb ~,
\end{align}
the problem \eqref{eq_problem_sip_gen} for fixed $\Lambdab$ can be expressed as
\begin{align} \label{eq_problem_sip_quadratic}
	\mathop{\mathrm{min}}\limits_{\ff} &~~	\ff^\hermit \QQ \ff - 2 \realp{ \pp^\hermit \ff }
	\\ \nonumber
	~~	\mathrm{s.t.} &~~ \abs{f_{n}} = 1 , \, \forall n ~.
\end{align}
The problem \eqref{eq_problem_sip_quadratic} can be solved using gradient projections iterations in \cite[Alg.~1]{analogBeamformerDesign_TSP_2017}.

\subsubsection{Optimize $\Lambdab$ for fixed $\FF$}
The subproblem of \eqref{eq_problem_sip_gen} for fixed $\FF$ is exactly the LS problem defined in \eqref{eq_ls_sip_prob}, whose solution is provided in \eqref{eq_ls_sip}.


The overall algorithm to solve \eqref{eq_problem_sip_gen} via alternating optimization of $\FF$ and $\Lambdab$ is summarized in Algorithm~\ref{alg_overall}.

\begin{algorithm}[t]
	\caption{ESPRIT-Oriented Precoder Design via Joint Optimization of $\FF$ and $\Lambdab$ in \eqref{eq_problem_sip_gen}}
	\label{alg_overall}
	\begin{algorithmic}[1]
	    \State \textbf{Input:} Baseline precoder $\FFopt$, transmit steering matrix $\AAb$, selection matrices $\JJ_1$ and $\JJ_2$, SIP error weight $\eta$, convergence threshold $\epsilon$.
	    \State \textbf{Output:} ESPRIT-oriented precoder $\FF$, SIP-restoring matrix $\Lambdab$.
	    \State \textbf{Initialization:}
	    \State Initialize the precoder as $\FF = \FFopt$.
	    \State Initialize the SIP-restoring matrix as $\Lambdab = \Lambdabhatls$ via \eqref{eq_ls_sip}.
	    \State \textbf{Alternating Optimization Iterations:} 
	    \State \textbf{repeat} 
	     \Indent
	     	\State Update $\FF$ in \eqref{eq_problem_sip_quadratic} via \cite[Alg.~1]{analogBeamformerDesign_TSP_2017}.
	        \State Update $\Lambdab$ via \eqref{eq_ls_sip}. 
	     \EndIndent
	     \State \textbf{until} the objective \eqref{eq_problem_sip} converges.
	\end{algorithmic}
	\normalsize
\end{algorithm}


\section{Simulation Results}\label{sec_sim}
To evaluate the performance of the proposed ESPRIT-oriented precoder design approach in Algorithm~\ref{alg_overall}, we perform numerical simulations using a mmWave setup with $\fc = 28 \, \rm{GHz}$, $\Ntx = 64$ and $d = \lambda /2$. The \ac{SNR} of the $\thn{\ell}$ path is defined as $\snrl = P \abs{\alphal}^2/\sigma^2$. In the following parts, we first present our approach for creating baseline precoders and provide illustrative examples on beampatterns associated to ESPRIT-oriented precoders to gain insights into how ESPRIT SIP considerations change the shape of the beampatterns. Then, we evaluate the \ac{AoD} estimation performance of the designed precoders.

\subsection{Baseline Construction via Codebook-Based Approach}\label{sec_baseline}
\begin{figure}
	\centering
	\includegraphics[width=0.9\linewidth]{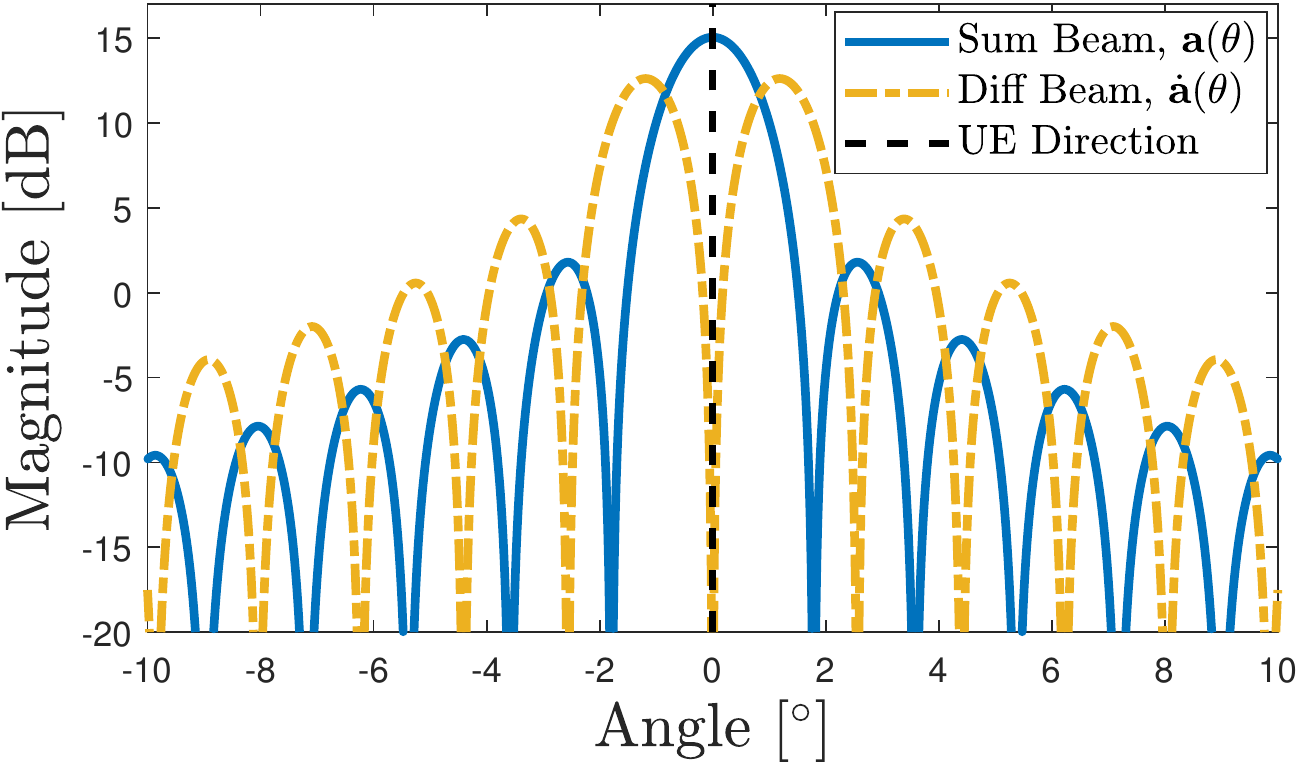}
	\vspace{-0.05in}
	\caption{Illustration of the beampatterns corresponding to the sum and difference beams in \eqref{eq_ffdig}, steered towards the \ac{AoD} $\theta = 0 \degree$. The sum beam provides the obvious benefit of maximizing the SNR towards the desired angle, while the difference beam improves \ac{AoD} accuracy in the small neighborhood around the targeted angle via its sharp curvature, which enables small angular deviations to induce large amplitude changes.}
	\label{fig_sum_diff_beam}
	\vspace{-0.1in}
\end{figure}

Following the idea in \cite{signalDesign_TVT_2022}, we propose to construct the baseline precoder $\FFopt$ via a codebook-based approach. Suppose that the \ac{BS} has a coarse \textit{a-priori} information on the  \acp{AoD} $\thetab$ in the form of uncertainty intervals, e.g., obtained via tracking routines \cite{mmwave_training_2016,Mendrzik_JSTSP_2019,signalDesign_TVT_2022}. 
Let $\mtu_{\ell} = [\thetaminl , \, \thetamaxl]$ denote the uncertainty interval for the \ac{AoD} of the $\thn{\ell}$ path and $\{ \theta_{\ell,i} \}_{i=1}^{\nl}$ the uniformly spaced \acp{AoD} covering $\mtul$, where the grid size $\nl$ is dictated by the $3 \, \rm{dB}$ beamwidth angular spacing \cite{zhang2018multibeam}. Accordingly, we define the codebook \cite{signalDesign_TVT_2022}
\begin{align} \label{eq_ffdig}
    \FFopt \triangleq \left[ \FFsum ~ \gamma \FFdiff \right] ~,
\end{align}
where 
\begin{align}\label{eq_codebook_def}
\FFsum &\triangleq \left[ \FFsum_0 \, \cdots \, \FFsum_{L-1} \right] ~, \\
\FFdiff &\triangleq \left[ \FFdiff_0 \, \cdots \, \FFdiff_{L-1} \right] ~,
\\
    \FFsum_{\ell} &\triangleq \left[ \atx^\conj(\theta_{\ell,1}) \, \cdots \, \atx^\conj(\theta_{\ell,\nl}) \right] ~, \\
    \FFdiff_{\ell} &\triangleq \left[ \atxdt^\conj(\theta_{\ell,1}) \, \cdots \, \atxdt^\conj(\theta_{\ell,\nl}) \right] ~,
\end{align}
for $\ell = 0, \ldots, L-1$, with $\atxdt(\theta) \triangleq {\partial \atx(\theta)}/{\partial \theta}$. Here, $\FFsum$ and $\FFdiff$ correspond to \textit{sum} (directional) and \textit{difference} (derivative) beams commonly employed in monopulse radar processing for accurate \ac{AoD} estimation \cite{monopulse_review}. Similar to radar, a combined use of these beams is shown to be optimal for positioning, as well \cite{signalDesign_TVT_2022}. In \eqref{eq_ffdig}, $\gamma$ represents the predefined weighting factor of the difference beams with respect to the sum beams, which is set to $\gamma = 0.01$ in simulations, and $\FFsum$ and $\FFdiff$ are normalized to have the same Frobenius norm before applying $\gamma$. To provide visualization and physical intuition, Fig.~\ref{fig_sum_diff_beam} shows the beampatterns of the sum and difference beams.

\subsection{Illustrative Examples for ESPRIT-Oriented Precoders}\label{sec_illust_ex}
\begin{figure}
        \begin{center}
        \vspace{-0.22in}
        \subfigure[]{
			 \label{fig_eta_1e0}
			 \includegraphics[width=0.43\textwidth]{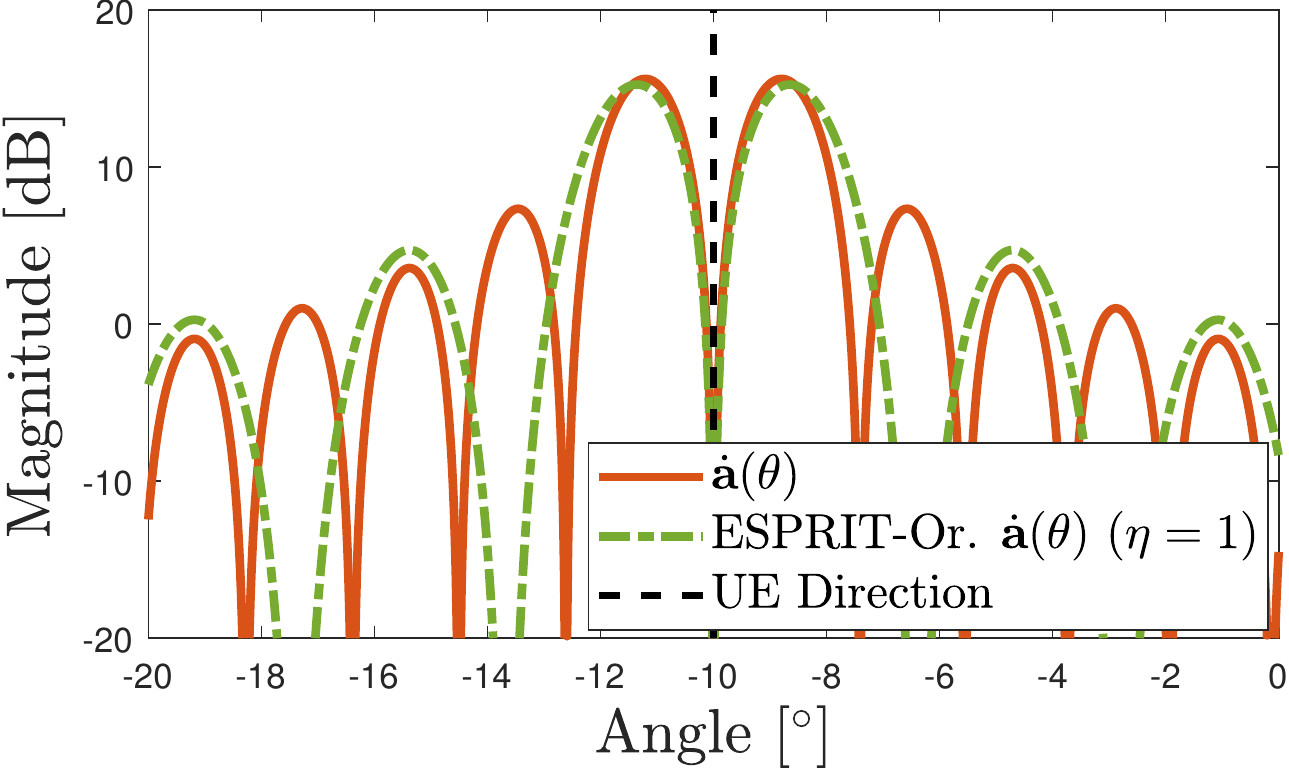}
		}
        \subfigure[]{
			 \label{fig_eta_1e4}
			 \includegraphics[width=0.43\textwidth]{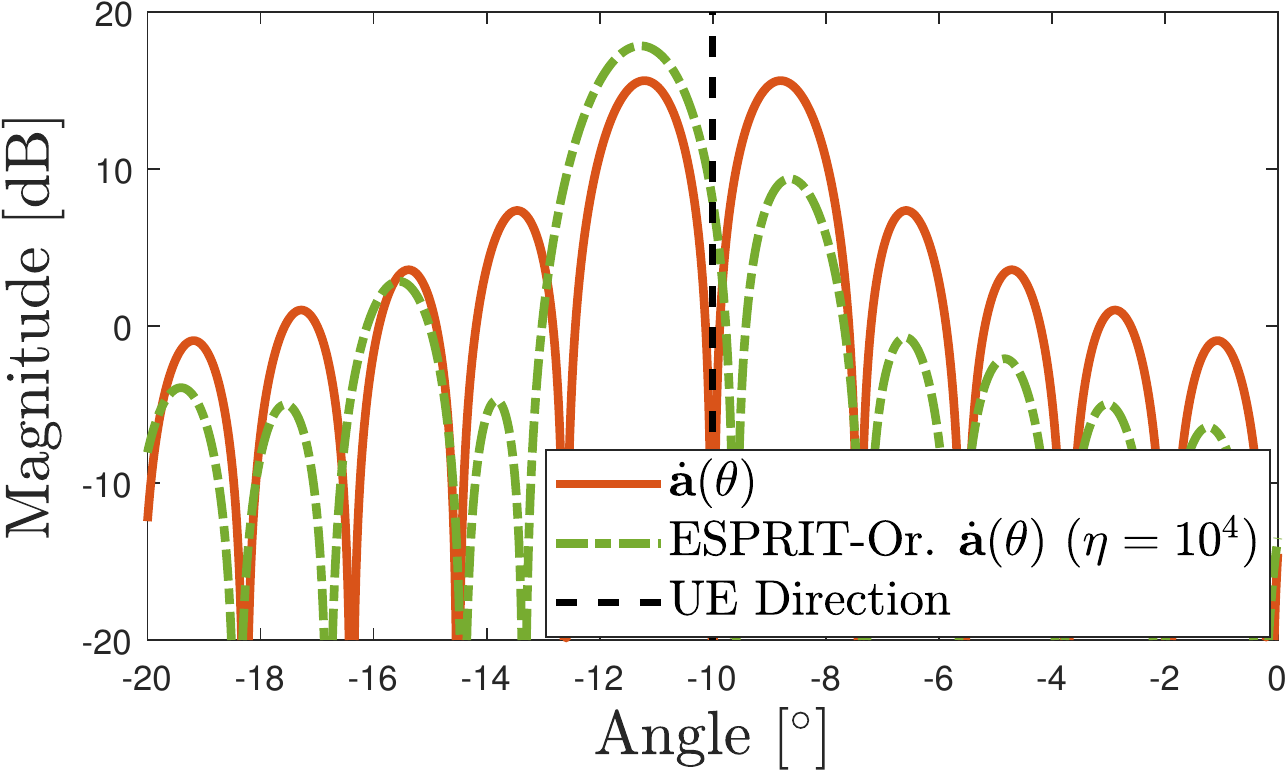}
		}
		
		\subfigure[]{
			 \label{fig_eta_1e6}
			 \includegraphics[width=0.43\textwidth]{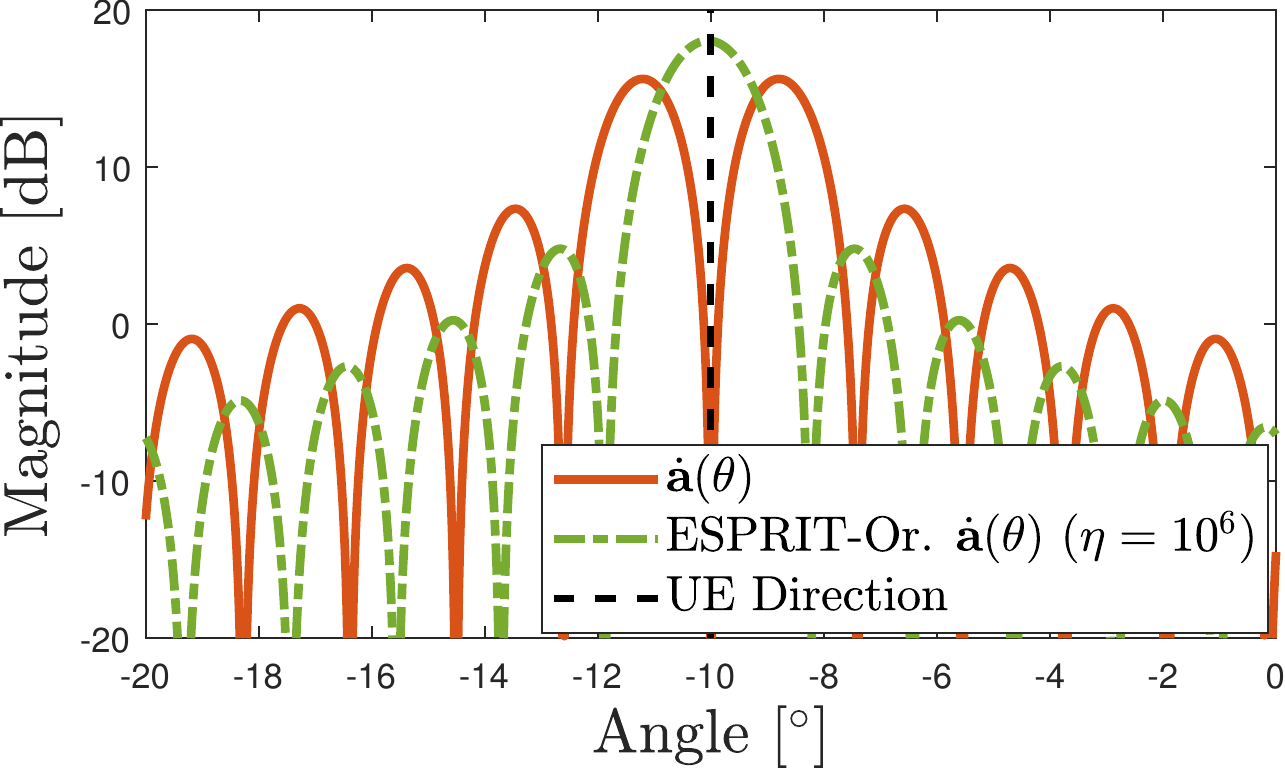}
		}
		
		\end{center}
		\vspace{-0.2in}
        \caption{The beampatterns of the ESPRIT-oriented precoders obtained via Algorithm~\ref{alg_overall} for varying $\eta$ values, where the baseline precoder $\FFopt$ is set to the difference beam with $\theta = -10 \degree$.}
        \label{fig_beampattern_eta}
        \vspace{-0.15in}
\end{figure}

\begin{figure}
	\centering
    \vspace{-0.2in}
	\includegraphics[width=0.9\linewidth]{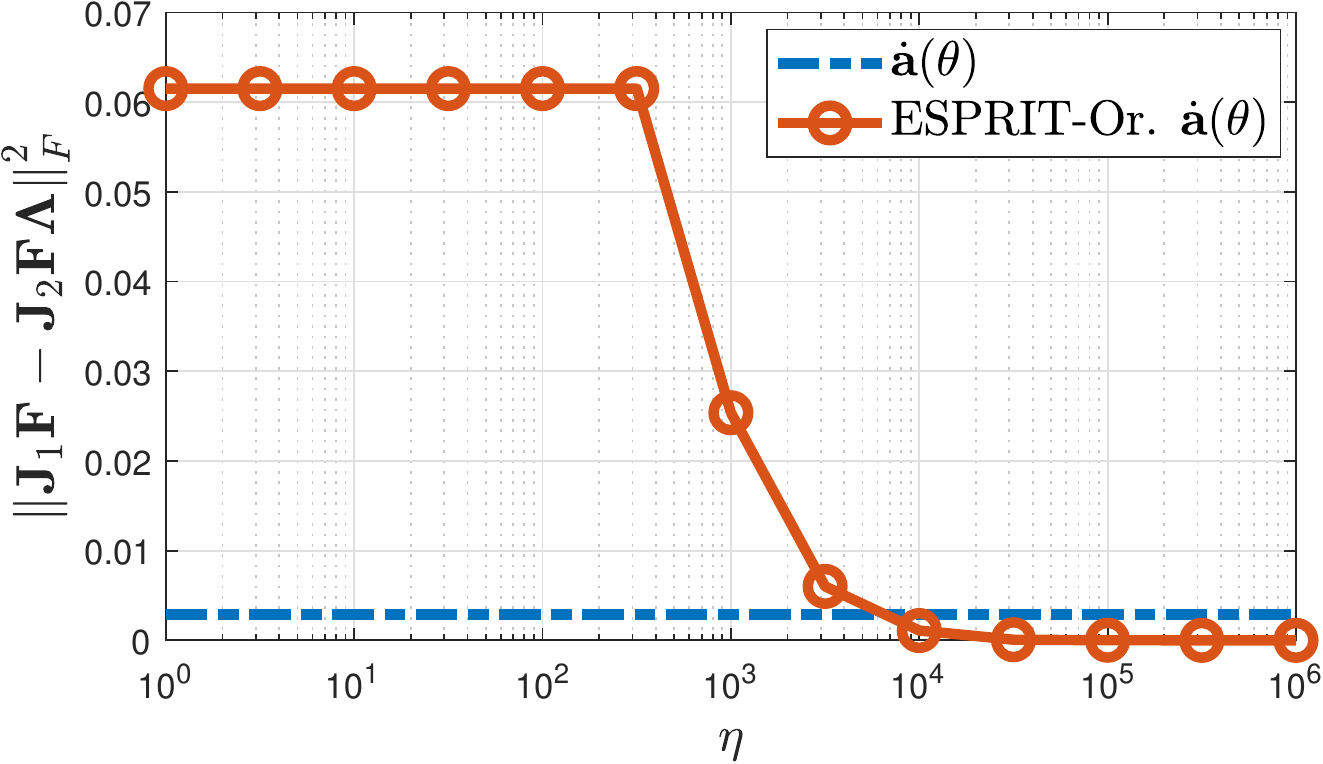}
	\vspace{-0.05in}
	\caption{SIP error in \eqref{eq_problem_sip} with respect to the penalty parameter $\eta$, where $\FFopt$ is the difference beam with $\theta = -10 \degree$.}
	\label{fig_sip_vs_eta}
	\vspace{-0.1in}
\end{figure}

Fig.~\ref{fig_beampattern_eta} shows the beampatterns of the ESPRIT-oriented precoders for three different $\eta$ values in Algorithm~\ref{alg_overall} by using the difference beam as the baseline. In Fig.~\ref{fig_sip_vs_eta}, the corresponding SIP errors in \eqref{eq_problem_sip} are plotted with respect to $\eta$. For small $\eta$, the ESPRIT-oriented precoder has a beampattern very close to that of the difference beam since Algorithm~\ref{alg_overall} places more emphasis on beampattern synthesis accuracy than on SIP approximation error, as seen from \eqref{eq_problem_sip}. As $\eta$ increases, SIP gains more emphasis, meaning that the resulting beam approaches the sum beam, for which the SIP is perfectly satisfied, as discussed in Sec.~\ref{sec_beamspace_esprit}. This leads us to the following important observation.

\textbf{Observation 1:} \textit{Phase-only ESPRIT-oriented precoder converges from \textbf{difference beam} towards \textbf{sum beam} as $\eta$ increases.}

\begin{figure}
	\centering
	\includegraphics[width=0.9\linewidth]{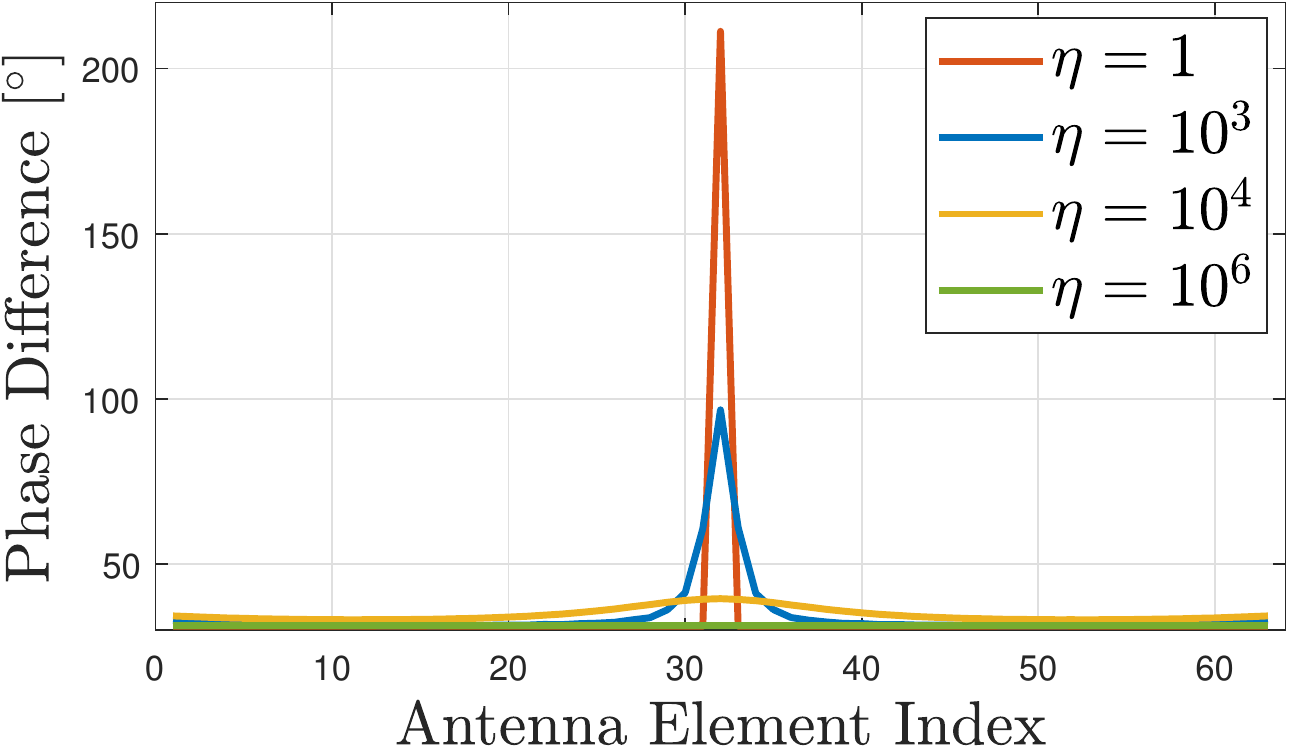}
	\caption{Phase changes of the ESPRIT-oriented precoder obtained via Algorithm~\ref{alg_overall} across the antenna elements for varying $\eta$ values, where the baseline precoder is taken as the difference beam with $\theta = -10 \degree$.}
	\label{fig_phase_vs_eta}
	\vspace{-0.15in}
\end{figure}

To provide further insights, we show in Fig.~\ref{fig_phase_vs_eta} the phase differences across the antenna elements of the ESPRIT-oriented precoder for various $\eta$ values. For small $\eta$, the ESPRIT-oriented precoder is close to the difference beam, which has a phase jump at the center of the array. The phase difference profile becomes more smooth as $\eta$ increases due to the SIP requirement, which causes the resulting beam to converge to the sum beam (which has uniform phase increments). Thus, the second important observation regarding ESPRIT-oriented precoders is stated as follows.

\textbf{Observation 2:} \textit{ESPRIT SIP requirement enforces \textbf{uniform phase increments} across antenna elements.}

\subsection{Evaluation of \ac{AoD} Estimation Performance}
To evaluate the \ac{AoD} estimation performance of the ESPRIT-oriented precoders designed via Algorithm~\ref{alg_overall}, we investigate the accuracy quantified through the \ac{RMSE} of $\thetab$, i.e.,
\begin{align}\label{eq_rmse}
    \RMSE_{\thetab} = \big( \E \big\{ \norms{\thetabhat - \thetab}^2 \big\} \big)^{1/2} ~,
\end{align}
where $\thetabhat = [\thetahat_0 \, \cdots \, \thetahat_{L-1}]^\mathsf{T}$ represents the estimate of $\thetab$ from $\yy$ in \eqref{eq_yy}. To obtain $\thetabhat$, we apply 1-D beamspace ESPRIT \cite{beamspace_ESPRIT} described in Sec.~\ref{sec_beamspace_esprit} on the observations $\yy$ in \eqref{eq_yy}. We run $100$ Monte Carlo trials with $50$ snapshots each to construct the covariance matrix for ESPRIT at each trial. The channel gains $\alphal$ are generated randomly across the snapshots by multiplying a fixed gain (determined based on $\snrl$) with a random zero-mean complex Gaussian coefficient with standard deviation $10$. In addition, based on the results in Sec.~\ref{sec_illust_ex}, we set $\eta = 10^5$ in Algorithm~\ref{alg_overall}. For performance benchmarking, we consider the following precoders:
\begin{itemize}
    \item \textit{Sum:} The precoder $\FFsum$ in \eqref{eq_codebook_def}, which by definition contains only unit-amplitude elements (i.e., steering vectors), leading to phase-only beamforming without further optimization.
    \item \textit{Sum-Diff:} The precoder $\FFopt$ in \eqref{eq_ffdig}, optimized to have unit-amplitude elements by using \cite[Alg.~1]{analogBeamformerDesign_TSP_2017}, which corresponds to a single $\FF$ update step in Algorithm~\ref{alg_overall}.
    \item \textit{Sum-Diff, ESPRIT-Or.:} The precoder obtained via the proposed ESPRIT-oriented precoder design algorithm in Algorithm~\ref{alg_overall}.
\end{itemize}
All the precoders are normalized to have the same Frobenius norm $\norm{\FF}_F$ so that the total transmit power in \eqref{eq_yy} remains the same among the different strategies for fair comparison. 

\begin{figure}
	\centering
	\includegraphics[width=0.9\linewidth]{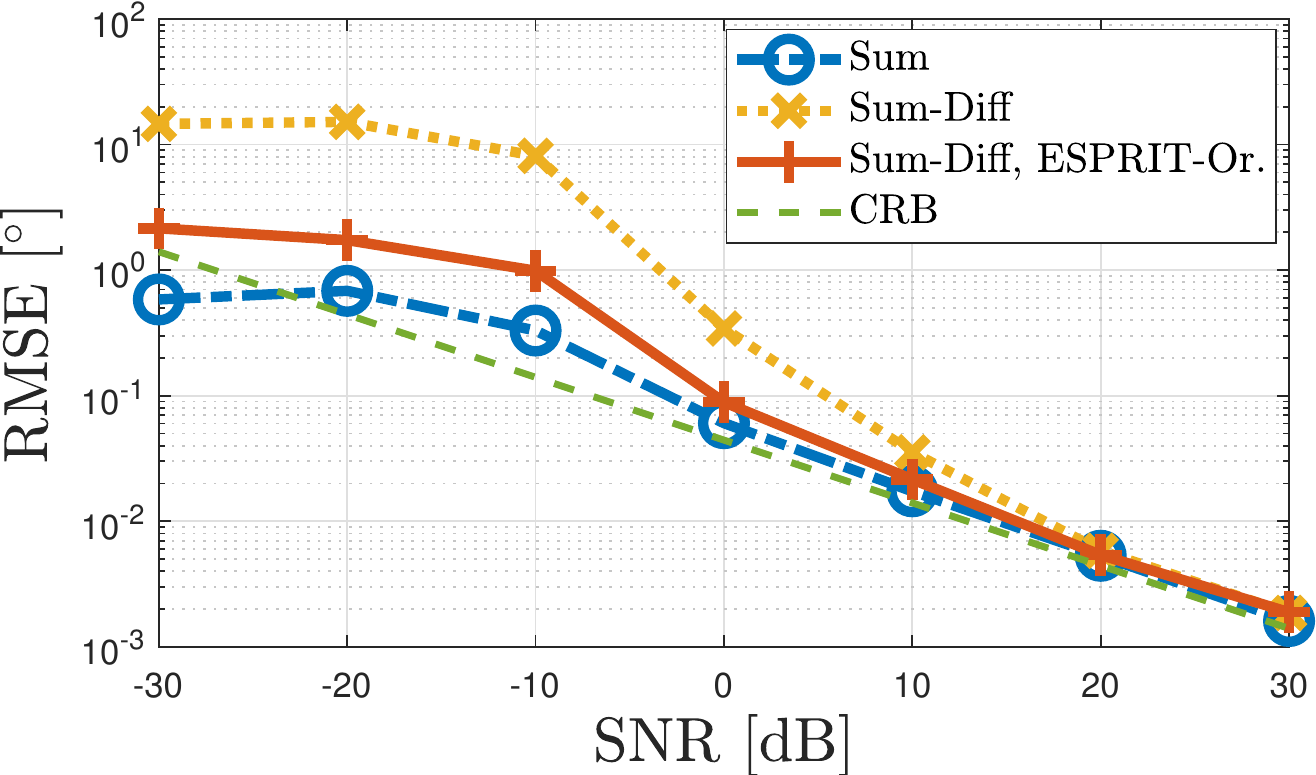}
	\caption{ESPRIT RMSEs obtained by the considered precoders with respect to SNR for a single-path scenario, where $\theta_0 = 20 \degree$ and $\mtu_0 = [17 \degree , \, 23 \degree]$.}
	\label{fig_rmse_snr_theta_20_delta_3_single_path}
\end{figure}

We first consider a single-path scenario with $\theta_0 = 20 \degree$ and $\mtu_0 = [17 \degree , \, 23 \degree]$. Fig.~\ref{fig_rmse_snr_theta_20_delta_3_single_path} shows the RMSEs obtained by the considered precoding strategies as a function of the SNR, also in comparison with the \ac{CRB}\footnote{Since the \acp{CRB} belonging to the different precoders are very close to each other, we only show the \ac{CRB} corresponding to $\FFsum$ for the sake of figure readability.}. It can be observed that the ESPRIT-oriented precoder provides noticeable improvement over the ESPRIT-unaware conventional sum-diff precoder at low \acp{SNR}, indicating the effectiveness of the proposed design strategy in Algorithm~\ref{alg_overall}. However, the conventional sum precoder outperforms the ESPRIT-oriented design at low \acp{SNR}, while the RMSEs of all the precoders converge to the CRB as the \ac{SNR} increases. This suggests that although Algorithm~\ref{alg_overall} succeeds in improving the performance, the sum precoder appears to be the best choice in this specific scenario. 

\begin{figure}
	\centering
	\includegraphics[width=0.9\linewidth]{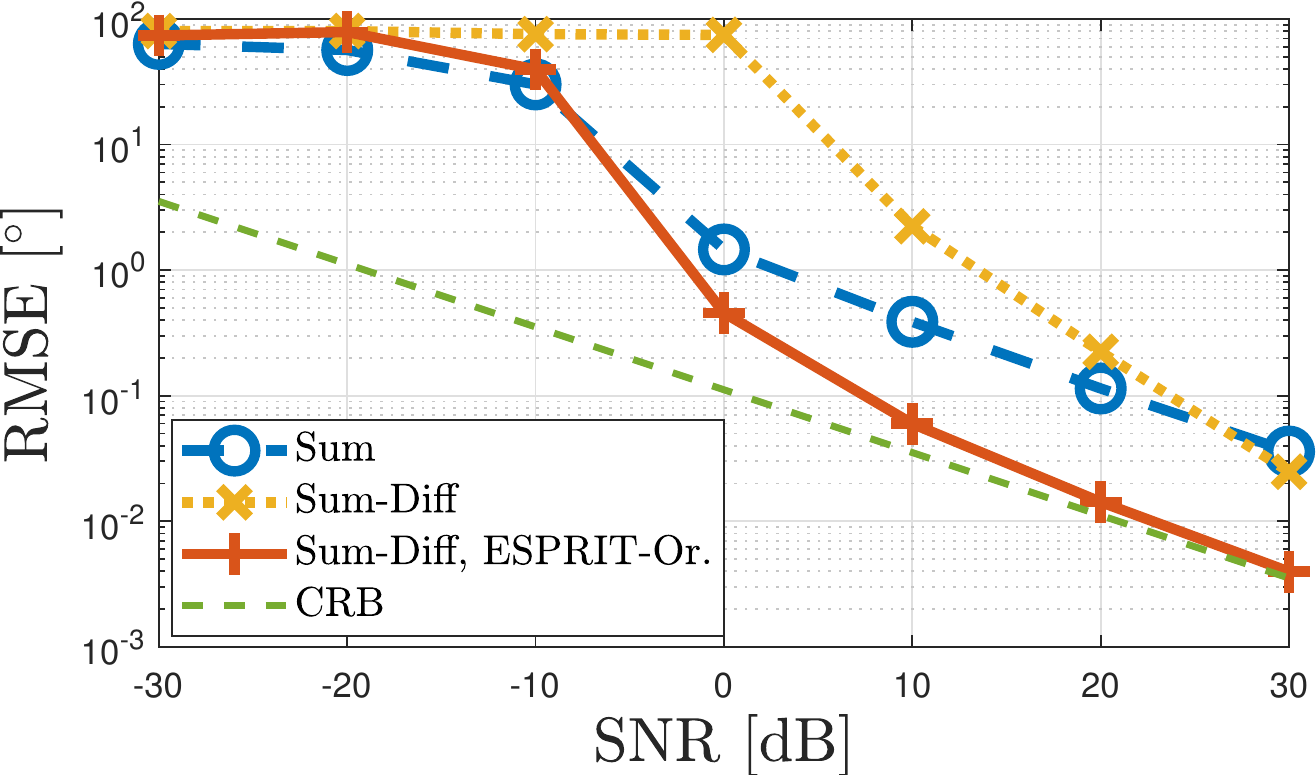}
	\caption{ESPRIT RMSEs obtained by the considered precoders with respect to SNR for a single-path scenario, where $\theta_0 = 70 \degree$ and $\mtu_0 = [67 \degree , \, 73 \degree]$.}
	\label{fig_rmse_snr_theta_70_delta_3_single_path}
	\vspace{-0.1in}
\end{figure}

\begin{figure}
        \begin{center}
        \subfigure[]{
			 \label{fig_unc_1}
			 \includegraphics[width=0.45\textwidth]{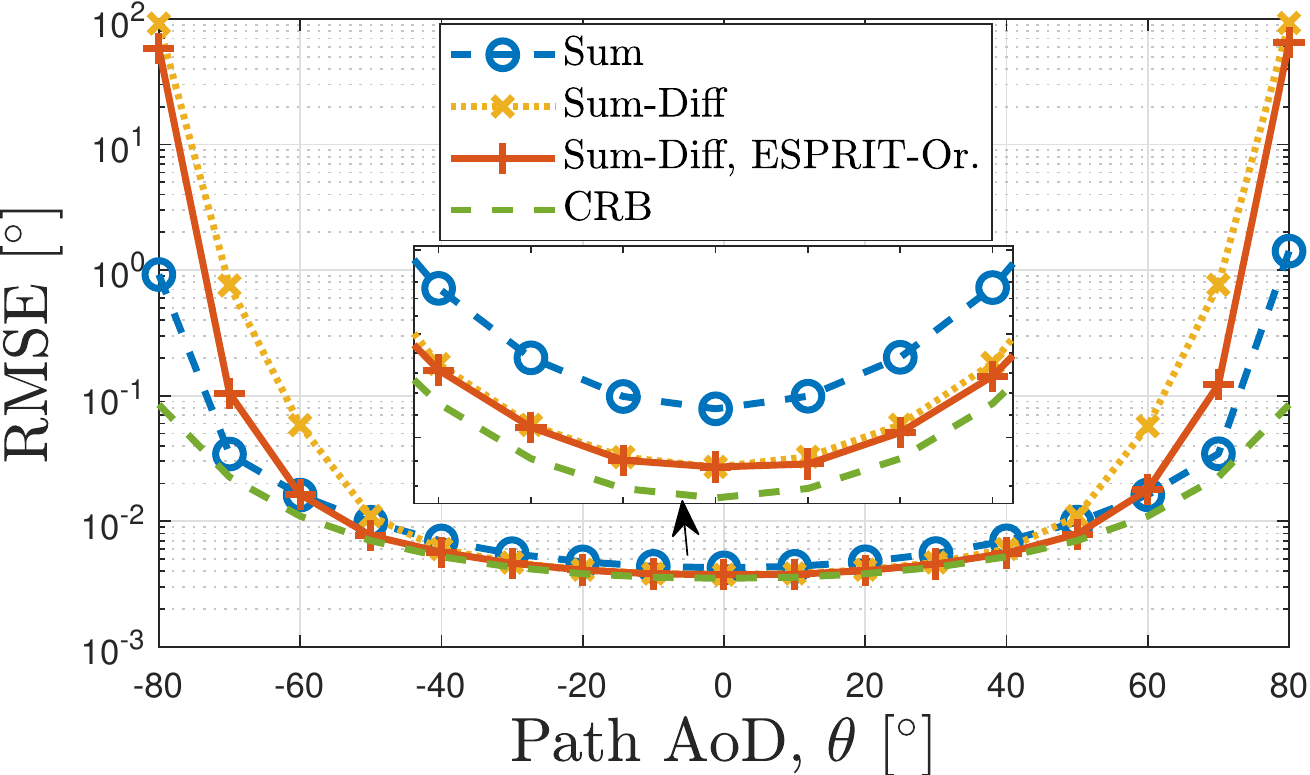}
		}
        \subfigure[]{
			 \label{fig_unc_3}
			 \includegraphics[width=0.45\textwidth]{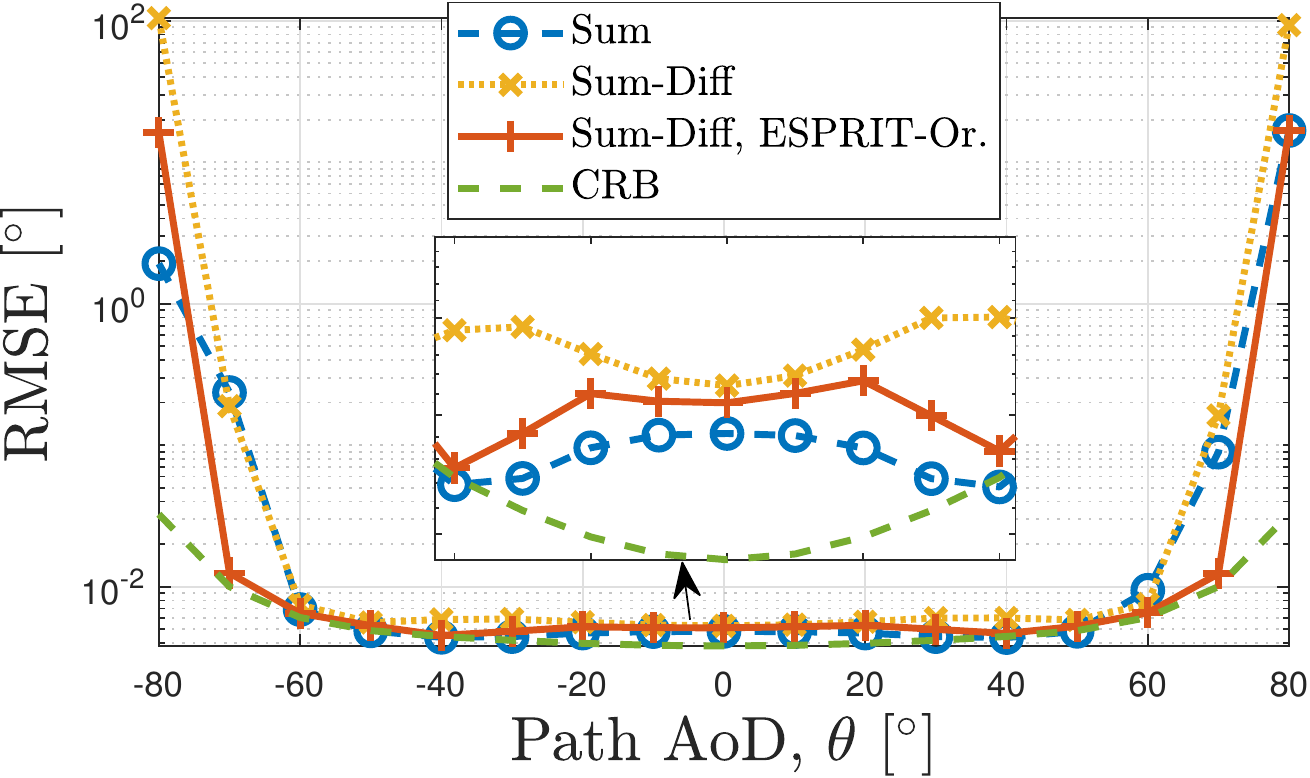}
		}
		
		\subfigure[]{
			 \label{fig_unc_5}
			 \includegraphics[width=0.45\textwidth]{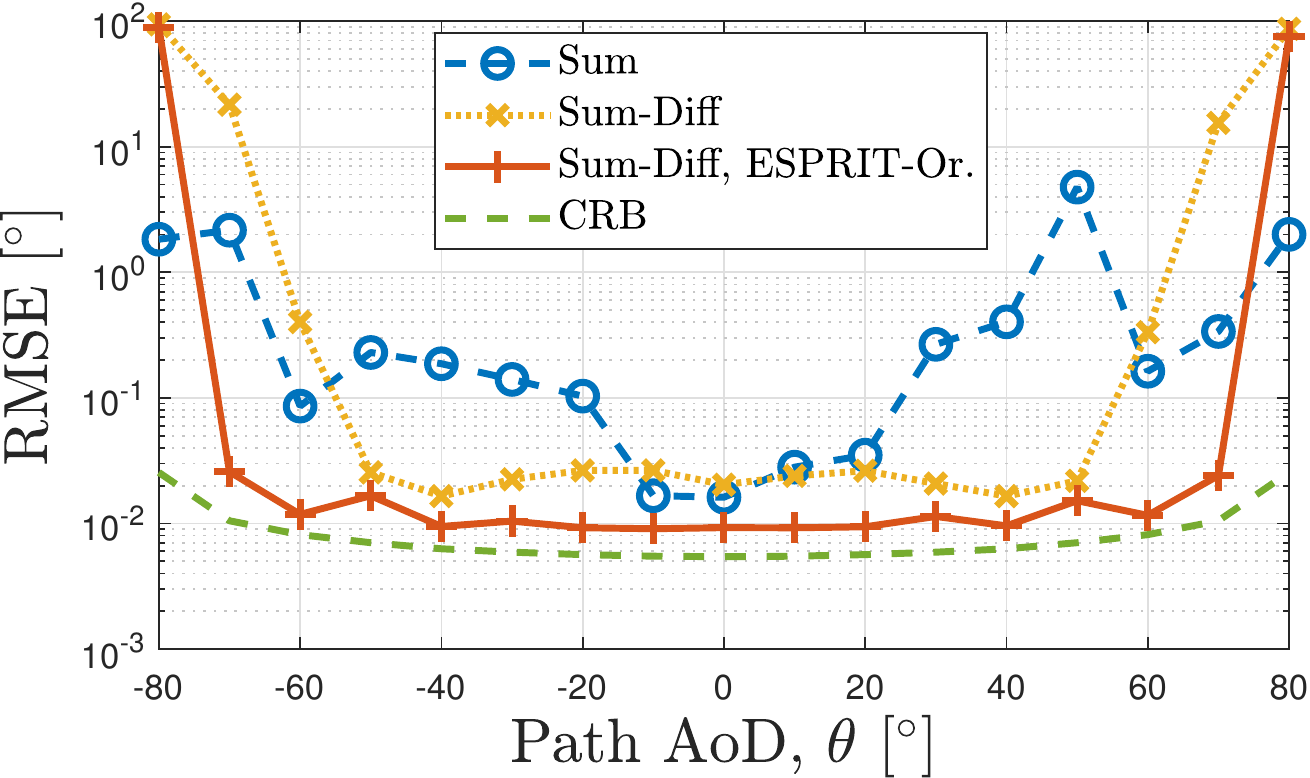}
		}
		
		\end{center}
		\vspace{-0.2in}
        \caption{ESPRIT RMSEs obtained by the considered precoders with respect to the path \ac{AoD} for a single-path scenario with \subref{fig_unc_1} $\pm 1$, \subref{fig_unc_3} $\pm 3$, and \subref{fig_unc_5} $\pm 5$ degrees of angular uncertainty for $\snr = 20 \, \rm{dB}$.}
        \label{fig_rmse_theta}
        \vspace{-0.2in}
\end{figure}

Next, we consider a different setting with $\theta_0 = 70 \degree$ and $\mtu_0 = [67 \degree , \, 73 \degree]$, whose results are reported in Fig.~\ref{fig_rmse_snr_theta_70_delta_3_single_path}. We observe that the proposed ESPRIT-oriented design significantly outperforms both the traditional sum precoder and sum-diff precoder in the medium and high SNR regimes, closing the gap to the \ac{CRB}. Comparing Fig.~\ref{fig_rmse_snr_theta_20_delta_3_single_path} and Fig.~\ref{fig_rmse_snr_theta_70_delta_3_single_path}, it is seen that performance gains provided by Algorithm~\ref{alg_overall} depend on the \ac{AoD} of the path. To further investigate this point, we plot in Fig.~\ref{fig_rmse_theta} the RMSE with respect to the path \ac{AoD} for a fixed SNR of $20 \, \rm{dB}$ with varying degrees of angular uncertainty. A common observation is that for all \acp{AoD}, the ESPRIT-oriented sum-diff precoder outperforms the standard sum-diff precoder, which does not consider the ESPRIT SIP conditions, suggesting that Algorithm~\ref{alg_overall} can provide considerable accuracy gains in ESPRIT-based estimation. For $\pm 1 \degree$ uncertainty, the ESPRIT-oriented precoder achieves lower RMSE than the sum precoder for $\theta \in [-60 \degree, 60 \degree]$ in agreement with \cite{signalDesign_TVT_2022}, while the trend becomes the opposite outside this interval. Looking at the $\pm 3 \degree$ uncertainty case, the sum precoder performs slightly better than the ESPRIT-oriented one around $\theta = 0 \degree$, while the latter can significantly outperform the former at the end-fire of the array, i.e., when the absolute value of the \ac{AoD} is above $60 \degree$. Furthermore, for the $\pm 5 \degree$ uncertainty case, the proposed ESPRIT-oriented design provides substantial gains over the sum precoder for almost the entire range of \ac{AoD} values, which further evidences the effectiveness of the proposed algorithm.

\begin{figure}
	\centering
    \vspace{-0.2in}
	\includegraphics[width=0.9\linewidth]{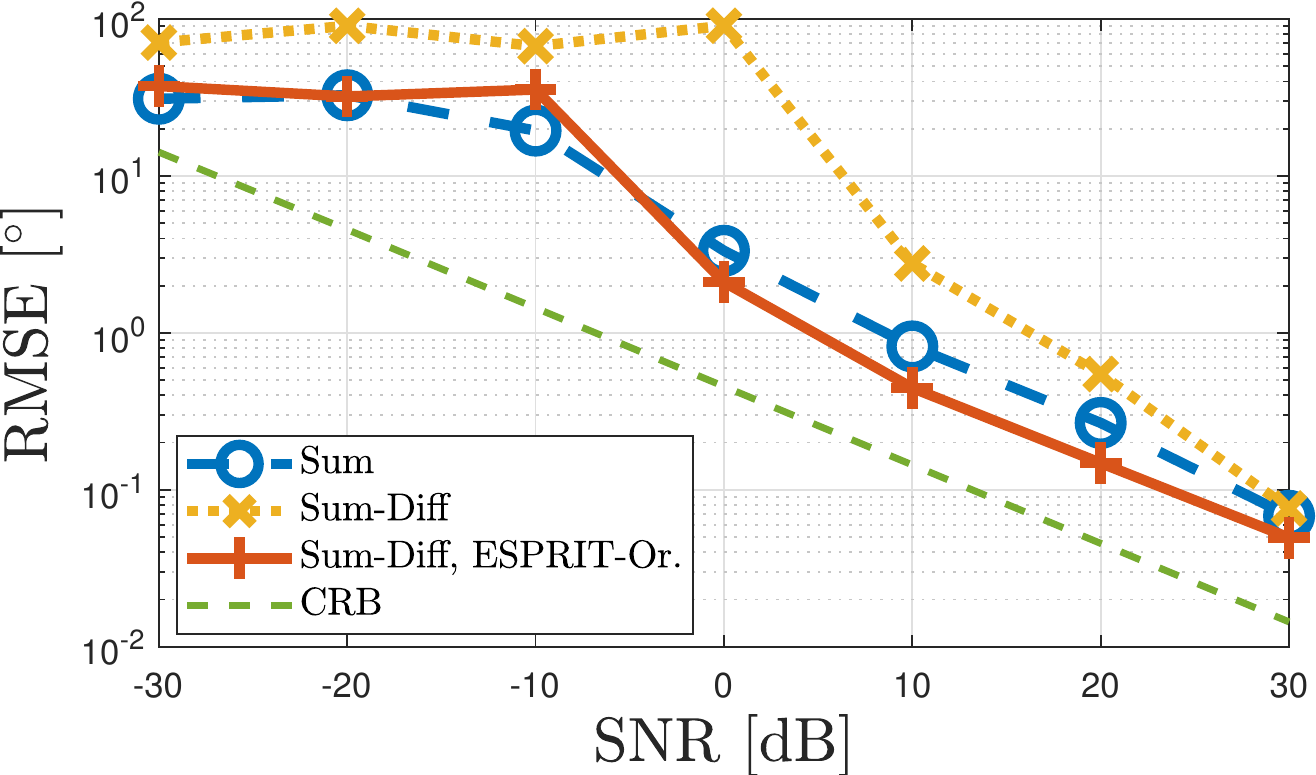}
	\vspace{-0.05in}
	\caption{ESPRIT RMSEs obtained by the considered precoders with respect to SNR for a two-path scenario, where $\thetab = [20\degree, 70\degree]$ with $\pm 3$ degrees of uncertainty for both paths.}
	\label{fig_rmse_snr_two_paths}
	\vspace{-0.1in}
\end{figure}

Finally, we investigate the RMSE performances for a two-path scenario with $\thetab = [20\degree, 70\degree]$, $\mtu_0 = [17 \degree , \, 23 \degree]$, $\mtu_1 = [67 \degree , \, 73 \degree]$, and $\snr = [20, 0] \, \rm{dB}$. Fig.~\ref{fig_rmse_snr_two_paths} plots the RMSE with respect to the SNR of the second path, where the SNRs of both paths are changed simultaneously while keeping their difference fixed. It is observed that the proposed ESPRIT-based design achieves higher accuracy than the benchmark schemes in the medium and high SNR regimes. The gap to the CRB can be attributed to the intrinsic suboptimality of ESPRIT \cite{esprit_tsp_91,doa_esprit_TAES_93} and to imperfect decorrelation of the paths in the estimated correlation matrix $\widetilde{\RRb}$.

\section{Concluding Remarks}
In this paper, we have studied the problem of mmWave precoder design tailored specifically to ESPRIT-based channel estimation. Considering the fact that standard precoders (i.e., sum beam) fail to achieve satisfactory performance in \ac{AoD} estimation and that \ac{CRB}-optimized precoders (sum-diff beam) destroy the \ac{SIP} of ESPRIT, leading to large degradations in ESPRIT accuracy, we have developed a novel ESPRIT-oriented precoder design approach that jointly optimizes the precoder and the \ac{SIP}-restoring matrix used in ESPRIT. Simulation results have provided valuable insights into how the \ac{SIP} requirement impacts the beampattern of the ESPRIT-oriented precoders and shown the effectiveness of the proposed design strategy. As future work, similar design principles can be employed to extend the current study to higher dimensions, i.e., 2-D uniform rectangular arrays (URAs) at both the \ac{BS} and the \ac{UE} sides, possibly with \ac{OFDM} transmission, leading to ESPRIT-oriented precoder and combiner designs for 5-D channel estimation (\ac{AoD}, \ac{AoA} and delay) \cite{Fan_ESPRIT_2021}.

\section*{Acknowledgment}
{
{This work was supported, in part, by the European Commission through the H2020 project Hexa-X (Grant Agreement no. 101015956), the MSCA-IF grant 888913 (OTFS-RADCOM), ICREA Academia Program, and Spanish R+D project PID2020-118984GB-I00.}}


\bibliographystyle{IEEEtran}
\bibliography{IEEEabrv,Sub/esprit_sip}

\begin{thebibliography}{10}
\providecommand{\url}[1]{#1}
\csname url@samestyle\endcsname
\providecommand{\newblock}{\relax}
\providecommand{\bibinfo}[2]{#2}
\providecommand{\BIBentrySTDinterwordspacing}{\spaceskip=0pt\relax}
\providecommand{\BIBentryALTinterwordstretchfactor}{4}
\providecommand{\BIBentryALTinterwordspacing}{\spaceskip=\fontdimen2\font plus
\BIBentryALTinterwordstretchfactor\fontdimen3\font minus
  \fontdimen4\font\relax}
\providecommand{\BIBforeignlanguage}[2]{{%
\expandafter\ifx\csname l@#1\endcsname\relax
\typeout{** WARNING: IEEEtran.bst: No hyphenation pattern has been}%
\typeout{** loaded for the language `#1'. Using the pattern for}%
\typeout{** the default language instead.}%
\else
\language=\csname l@#1\endcsname
\fi
#2}}
\providecommand{\BIBdecl}{\relax}
\BIBdecl

\bibitem{b5g_commag_2021}
S.~Bartoletti \emph{et~al.}, ``Positioning and sensing for vehicular safety
  applications in {5G} and beyond,'' \emph{IEEE Communications Magazine},
  vol.~59, no.~11, pp. 15--21, 2021.

\bibitem{dwivedi2021positioning}
S.~Dwivedi \emph{et~al.}, ``Positioning in {5G} networks,'' \emph{IEEE
  Communications Magazine}, vol.~59, no.~11, pp. 38--44, 2021.

\bibitem{Fascista_ICASSP2020}
A.~Fascista \emph{et~al.}, ``Low-complexity accurate mmwave positioning for
  single-antenna users based on angle-of-departure and adaptive beamforming,''
  in \emph{IEEE International Conference on Acoustics, Speech and Signal
  Processing ({ICASSP})}, 2020, pp. 4866--4870.

\bibitem{signalDesign_TVT_2022}
M.~F. Keskin \emph{et~al.}, ``Optimal spatial signal design for mmwave
  positioning under imperfect synchronization,'' \emph{IEEE Transactions on
  Vehicular Technology}, vol.~71, no.~5, pp. 5558--5563, 2022.

\bibitem{TR38.855}
3rd Generation Partnership Project~(3GPP), ``Study on {NR} positioning support
  {TR} 38.855,'' \emph{Technical Specification Group Radio Access Network},
  2019.

\bibitem{Fascista_WCL}
A.~Fascista \emph{et~al.}, ``Low-complexity downlink channel estimation in
  mmwave multiple-input single-output systems,'' \emph{IEEE Wireless
  Communications Letters}, vol.~11, no.~3, pp. 518--522, 2022.

\bibitem{overviewISAC_2021}
J.~A. Zhang \emph{et~al.}, ``An overview of signal processing techniques for
  joint communication and radar sensing,'' \emph{IEEE Journal of Selected
  Topics in Signal Processing}, vol.~15, no.~6, pp. 1295--1315, 2021.

\bibitem{tsai2018millimeter}
Y.~Tsai \emph{et~al.}, ``Millimeter-wave beamformed full-dimensional mimo
  channel estimation based on atomic norm minimization,'' \emph{IEEE
  Transactions on Communications}, vol.~66, no.~12, pp. 6150--6163, 2018.

\bibitem{lee2016channel}
J.~Lee \emph{et~al.}, ``Channel estimation via orthogonal matching pursuit for
  hybrid mimo systems in millimeter wave communications,'' \emph{IEEE
  Transactions on Communications}, vol.~64, no.~6, pp. 2370--2386, 2016.

\bibitem{alkhateeb2014channel}
A.~Alkhateeb \emph{et~al.}, ``Channel estimation and hybrid precoding for
  millimeter wave cellular systems,'' \emph{IEEE journal of selected topics in
  signal processing}, vol.~8, no.~5, pp. 831--846, 2014.

\bibitem{JiaGeZhuWym21}
F.~Jiang \emph{et~al.}, ``High-dimensional channel estimation for simultaneous
  localization and communications,'' in \emph{2021 IEEE WCNC}, Nanjing, China,
  2021.

\bibitem{WenKulWitWym19}
F.~{Wen} \emph{et~al.}, ``{5G} positioning and mapping with diffuse
  multipath,'' \emph{IEEE Transactions on Wireless Communications}, vol.~20,
  no.~2, pp. 1164--1174, 2021.

\bibitem{gridless_ESPRIT_JSTSP_2021}
J.~Zhang \emph{et~al.}, ``Gridless channel estimation for hybrid mmwave {MIMO}
  systems via tensor-{ESPRIT} algorithms in {DFT} beamspace,'' \emph{IEEE
  Journal of Selected Topics in Signal Processing}, vol.~15, no.~3, pp.
  816--831, 2021.

\bibitem{Fan_ESPRIT_2021}
\BIBentryALTinterwordspacing
F.~Jiang \emph{et~al.}, ``Beamspace multidimensional {ESPRIT} approaches for
  simultaneous localization and communications,'' 2021. [Online]. Available:
  \url{https://arxiv.org/abs/2111.07450}
\BIBentrySTDinterwordspacing

\bibitem{WenGarKulWitWym18}
F.~Wen \emph{et~al.}, ``Tensor decomposition based beamspace {ESPRIT} for
  millimeter wave mimo channel estimation,'' in \emph{IEEE GLOBECOM}, Abu
  Dhabi, United Arab Emirates, 2018.

\bibitem{beamspace_ESPRIT}
G.~Xu \emph{et~al.}, ``Beamspace {ESPRIT},'' \emph{IEEE Transactions on Signal
  Processing}, vol.~42, no.~2, pp. 349--356, 1994.

\bibitem{precoderNil2018}
N.~{Garcia} \emph{et~al.}, ``Optimal precoders for tracking the {AoD} and {AoA}
  of a {mmWave} path,'' \emph{IEEE Transactions on Signal Processing}, vol.~66,
  no.~21, pp. 5718--5729, Nov 2018.

\bibitem{Fascista_RIS}
A.~Fascista \emph{et~al.}, ``{RIS}-aided joint localization and synchronization
  with a single-antenna receiver: Beamforming design and low-complexity
  estimation,'' \emph{IEEE Journal of Selected Topics in Signal Processing},
  vol.~16, no.~5, pp. 1141--1156, 2022.

\bibitem{mmWave_Tracking_2020_TVT}
D.~Zhang \emph{et~al.}, ``Beam allocation for millimeter-wave {MIMO} tracking
  systems,'' \emph{IEEE Transactions on Vehicular Technology}, vol.~69, no.~2,
  pp. 1595--1611, 2020.

\bibitem{mmWave_Tracking_2020_TCOM}
Y.~Yang \emph{et~al.}, ``Bayesian beamforming for mobile millimeter wave
  channel tracking in the presence of {DOA} uncertainty,'' \emph{IEEE
  Transactions on Communications}, vol.~68, no.~12, pp. 7547--7562, 2020.

\bibitem{RoyKai89}
R.~Roy \emph{et~al.}, ``{ESPRIT}-estimation of signal parameters via rotational
  invariance techniques,'' \emph{IEEE Transactions on Acoustics, Speech, and
  Signal Processing}, vol.~37, no.~7, pp. 984--995, Jul. 1989.

\bibitem{analogBeamformerDesign_TSP_2017}
J.~Tranter \emph{et~al.}, ``Fast unit-modulus least squares with applications
  in beamforming,'' \emph{IEEE Transactions on Signal Processing}, vol.~65,
  no.~11, pp. 2875--2887, 2017.

\bibitem{phasedArray_2016}
S.~H. Talisa \emph{et~al.}, ``Benefits of digital phased array radars,''
  \emph{Proceedings of the IEEE}, vol. 104, no.~3, pp. 530--543, 2016.

\bibitem{mmwave_training_2016}
J.~C. Aviles \emph{et~al.}, ``Position-aided mm-wave beam training under {NLOS}
  conditions,'' \emph{IEEE Access}, vol.~4, pp. 8703--8714, 2016.

\bibitem{Mendrzik_JSTSP_2019}
R.~Mendrzik \emph{et~al.}, ``Enabling situational awareness in millimeter wave
  massive {MIMO} systems,'' \emph{IEEE Journal of Selected Topics in Signal
  Processing}, vol.~13, no.~5, pp. 1196--1211, 2019.

\bibitem{zhang2018multibeam}
J.~A. Zhang \emph{et~al.}, ``Multibeam for joint communication and radar
  sensing using steerable analog antenna arrays,'' \emph{IEEE Transactions on
  Vehicular Technology}, vol.~68, no.~1, pp. 671--685, 2018.

\bibitem{monopulse_review}
U.~Nickel, ``Overview of generalized monopulse estimation,'' \emph{IEEE
  Aerospace and Electronic Systems Magazine}, vol.~21, no.~6, pp. 27--56, 2006.

\bibitem{esprit_tsp_91}
B.~Ottersten \emph{et~al.}, ``Performance analysis of the total least squares
  {ESPRIT} algorithm,'' \emph{IEEE Transactions on Signal Processing}, vol.~39,
  no.~5, pp. 1122--1135, 1991.

\bibitem{doa_esprit_TAES_93}
F.~Li \emph{et~al.}, ``Performance analysis for {DOA} estimation algorithms:
  unification, simplification, and observations,'' \emph{IEEE Transactions on
  Aerospace and Electronic Systems}, vol.~29, no.~4, pp. 1170--1184, 1993.

\end{thebibliography}

\end{document}